\documentclass{webofc}
\usepackage[varg]{txfonts}   

\usepackage{overpic}
\usepackage{xcolor}
\usepackage{tikz}
\usetikzlibrary{patterns}
\usetikzlibrary{plotmarks}

\definecolor{colorArSc}{RGB}{255,102,51}
\definecolor{colorPP}{RGB}{0,0,204}
\definecolor{colorNN}{RGB}{0,0,204}
\definecolor{colorNNWorld}{RGB}{153,153,255}
\definecolor{colorBeBe}{RGB}{0,153,0}
\definecolor{colorXeLa}{RGB}{204,0,255}
\definecolor{colorPbPb}{RGB}{204,0,0}
\definecolor{colorAuAu}{RGB}{204,0,0}

\begin{document}
\title{Highlights from the NA61/SHINE strong-interactions\\
programme}

\author{\firstname{Magdalena} \lastname{Kuich}\inst{1}\fnsep\thanks{\email{magdalena.kuich@cern.ch}} for the NA61/SHINE Collaboration
}

\institute{Faculty of Physics, University of Warsaw, Pasteura 5, 02-093 Warsaw, Poland
          }

\abstract{
  NA61/SHINE is a multipurpose fixed-target facility at the CERN SPS. The main goals of the NA61/SHINE strong-interactions programme are to discover the critical point of strongly interacting matter as well as to study the properties of the onset of deconfinement. In order to reach these goals, a study of hadron production properties is performed in nucleus-nucleus, proton-proton and proton-nucleus interactions as a function of collision energy and size of the colliding nuclei. In this contribution, the NA61/SHINE results from a strong interaction measurement programme are presented. In particular, the latest results from different reactions \textit{p+p}, Be+Be, Ar+Sc, and Pb+Pb on hadron spectra, as well as intermittency, higher-order moments of multiplicity fluctuations and spectator induced electromagnetic effects are discussed. 
}
\maketitle
\section{NA61/SHINE experiment}
\label{intro}
NA61/SHINE detection system is a large acceptance hadron spectrometer based on a set of eight Time Projection Chambers complemented by Time-of-Flight detectors. This setup allows for precise momentum reconstruction and identification of charged particles. The high-resolution forward calorimeter, the Projectile Spectator Detector (PSD), measures energy flow around the beam direction, so-called Forward Energy ($E_F$), which in nucleus-nucleus 
can be related to the collision centrality~\cite{facilitypaper}. 

NA61/SHINE performed a two-dimensional scan in collision energy (13A-150A GeV/c and system size (p+p, p+Pb, Be+Be, Ar+Sc, Xe+La, Pb+Pb) to study the phase diagram of strongly interacting matter.
Currently, NA61/SHINE pursues the physics programme focused on the open charm production measurements in order to determine the mechanism of open charm production, the impact of the onset of deconfinement on open charm production and investigate the impact of the QGP formation on $J/\Psi$ production.

\section{Study of the onset of deconfinement and onset of fireball}
\label{sec:0D_OF}
\subsection{Particle production properties}
The main motivation of the NA61/SHINE physics programme are the predictions of the Statistical Model of the Early Stage (SMES)~\cite{smes}. The theory predicts a 1$^\textup{st}$ order phase transition from the QGP to a hadron matter phase between top AGS and top SPS energies. In the transition region, constant temperature and pressure in the mixed-phase and an increase of the number of internal degrees of freedom is expected. SMES proposes several observables, which indicates the phase transition.

\begin{figure}[h]
\centering
\includegraphics[width=0.38\textwidth,clip]{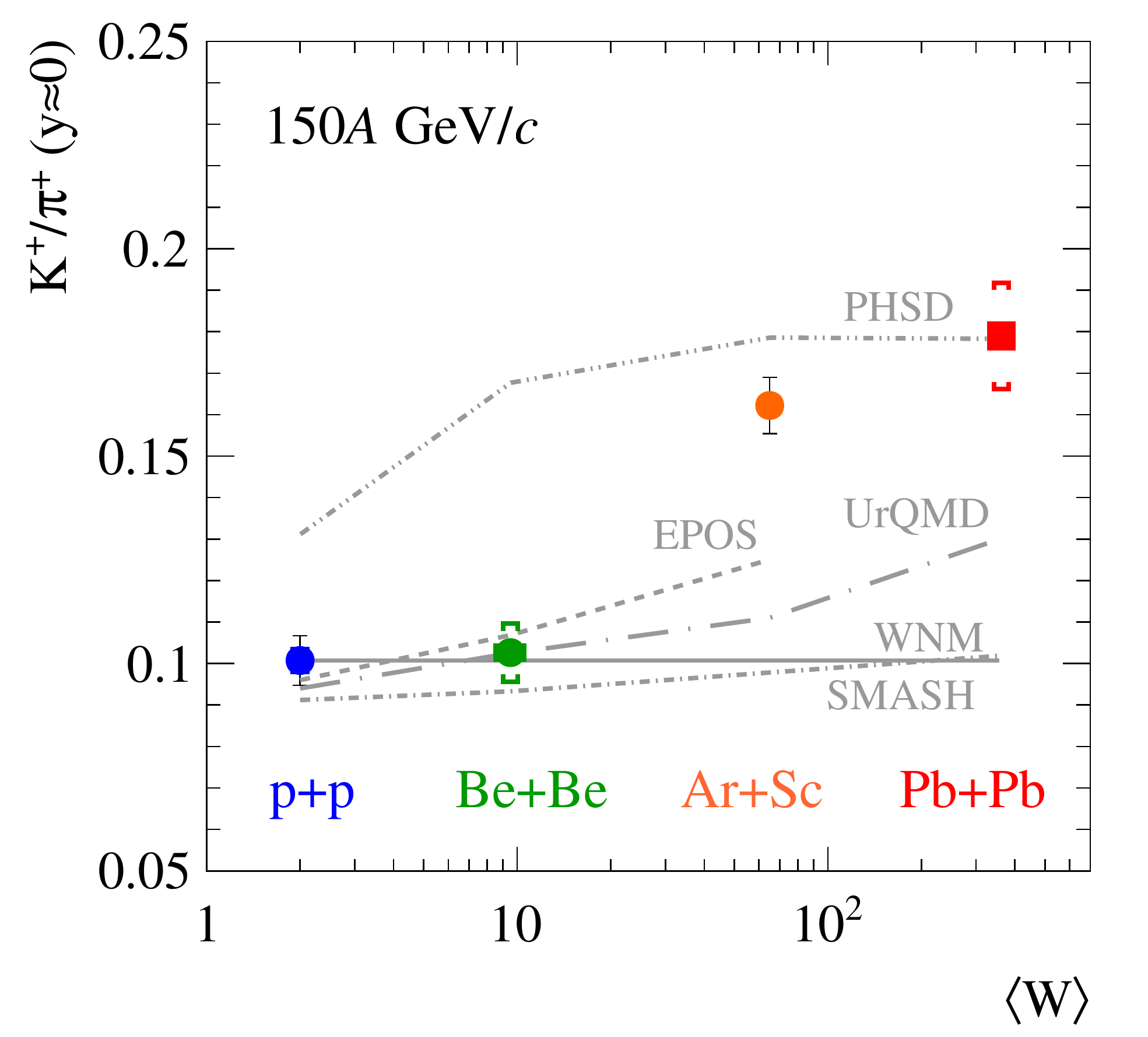}
\includegraphics[width=0.38\textwidth,clip]{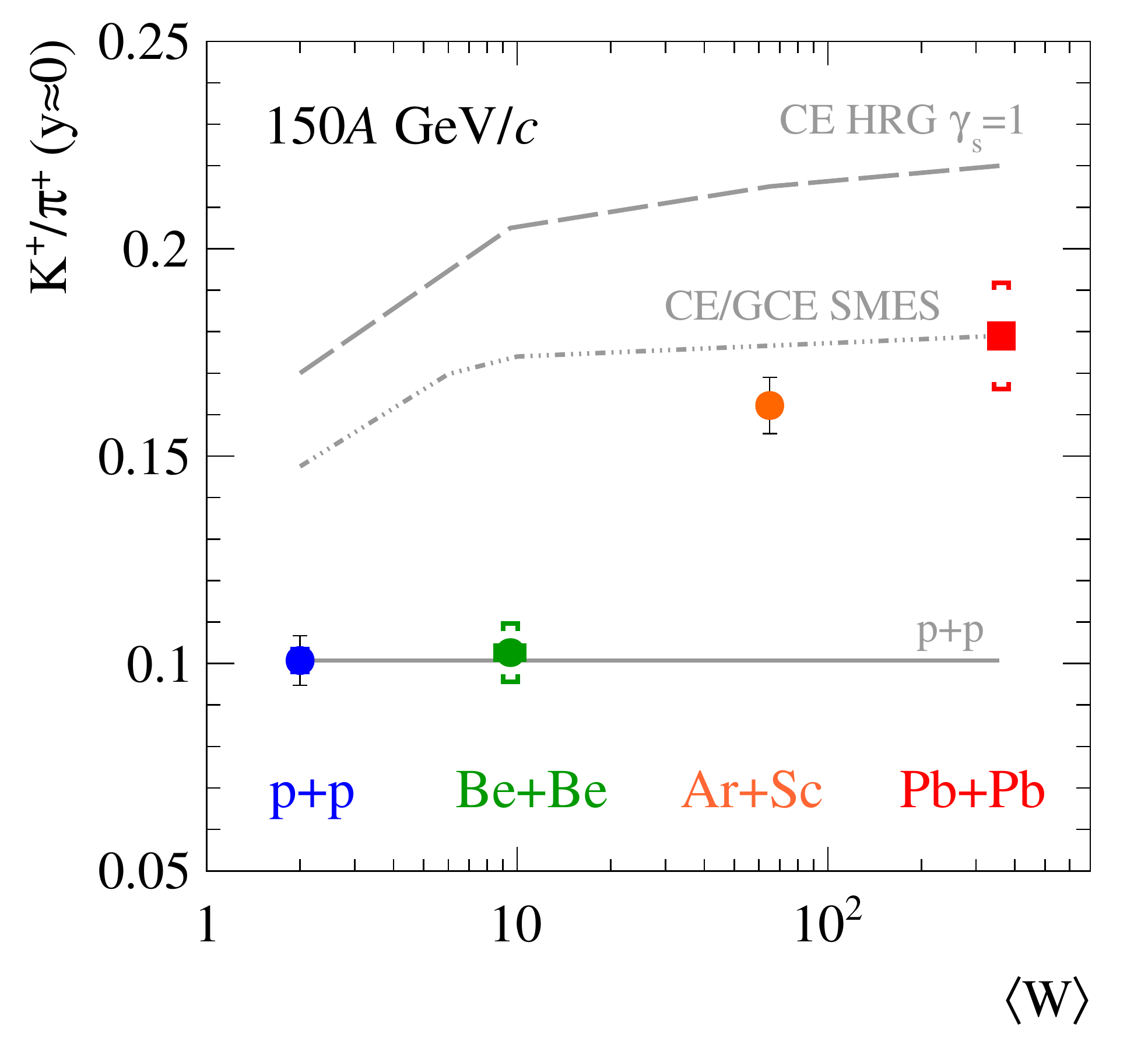}
\caption{System size dependence of the $K^+/\pi^+$ ratio obtained at beam momenta of 150$A$ GeV/$c$ ($\sqrt{s_{NN}}\approx17$ GeV) compared with dynamical (\textit{left}) and statistical (\textit{right}) models.}
\label{fig:onset_vs_models}
\end{figure}

Fig.~\ref{fig:onset_vs_models} presents the system-size dependence of $K^+/\pi^+$ ratio at mid-rapidity obtained by NA61/SHINE: \textit{p+p}~\cite{pp_identified}, Be+Be~\cite{BeBe_idenified} and Ar+Sc (preliminary); and NA49: Pb+Pb~\cite{PbPb_identified}; at 150$A$/158$A$ GeV/$c$ beam momenta. 
The $K^+/\pi^+$ ratio was predicted within SMES as one of the signatures of the onset of deconfinement. The results were compared with few dynamical models. Those without phase transition (EPOS~\cite{epos2}, UrQMD~\cite{hrg_urqmd} and SMASH~\cite{smash}) agree with the results from small systems (\textit{p+p} and Be+Be), while fail to describe the results from heavier systems (Ar+Sc and Pb+Pb). In contrast to the one with phase transition (PHSD~\cite{phsd1,phsd2}), which follows the trend determined by the heaviest system (Pb+Pb) and overestimates the ratio for smaller systems. Similarly, testes statistical models: with (renormalised SMES~\cite{smes_extended}) and without (HRG~\cite{hrg_urqmd}) the phase transition tends to overestimate the $K^+/\pi^+$ ratio especially in the small-mass region of colliding nuclei.

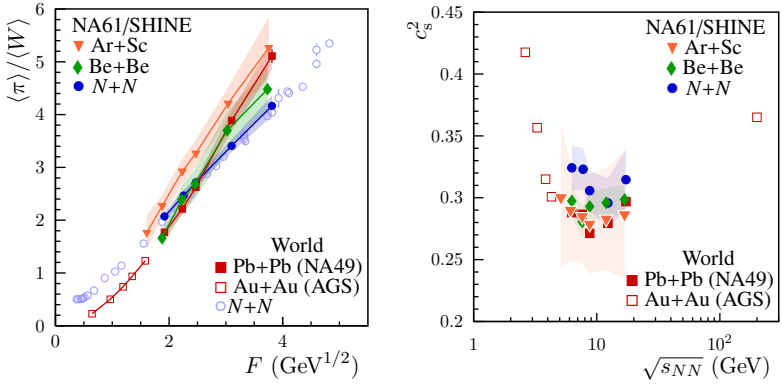
\begin{figure}[h]
\centering
\begin{overpic}[width=0.4\textwidth,clip]{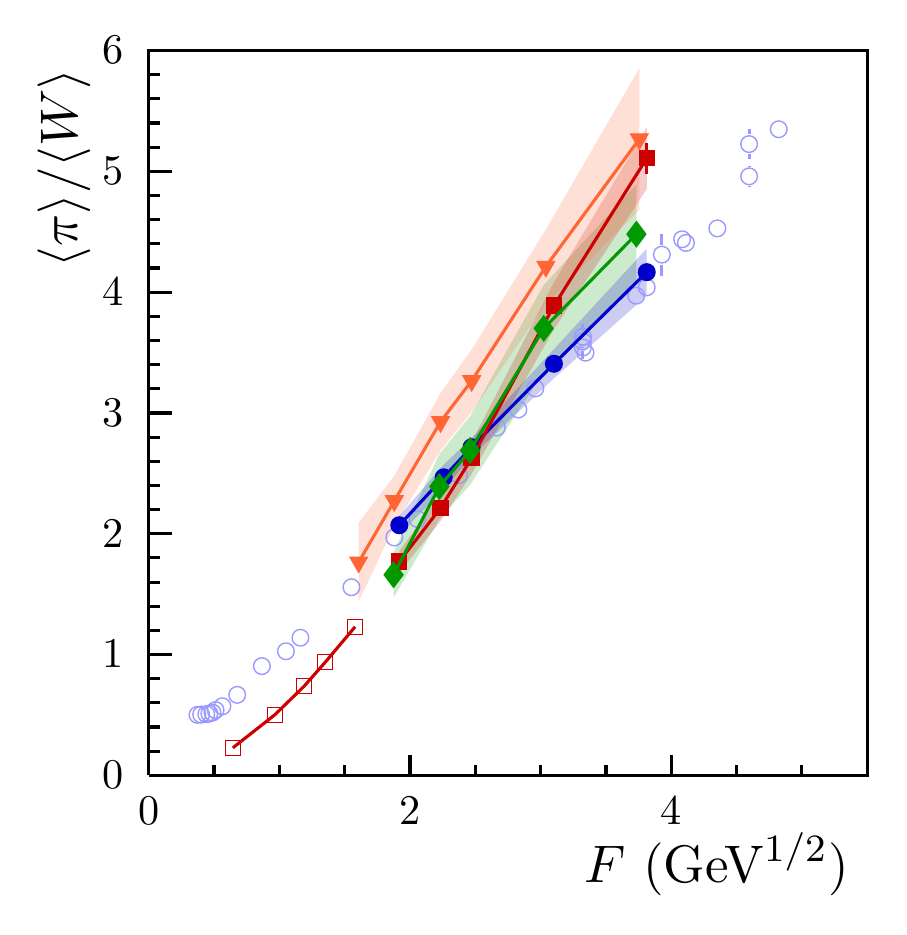}
 \put(20,89){\footnotesize NA61/SHINE}
 \put(20,84){\footnotesize \textcolor{colorArSc}{$\blacktriangledown$} Ar+Sc}
 \put(20,79){\footnotesize \textcolor{colorBeBe}{$\blacklozenge$} Be+Be}
 \put(20,74){\textcolor{colorNN}{$\bullet$} {\footnotesize $N$+$N$}}
 \put(69,35){\footnotesize World}
 \put(55,30){\footnotesize \textcolor{colorPbPb}{$\blacksquare$} Pb+Pb (NA49)}
 \put(55,25){\footnotesize \textcolor{colorAuAu}{$\square$} Au+Au (AGS)}
 \put(55,20){\textcolor{colorNNWorld}{$\circ$}{\footnotesize $N$+$N$}}
\end{overpic}
\begin{overpic}[width=0.41\textwidth,clip]{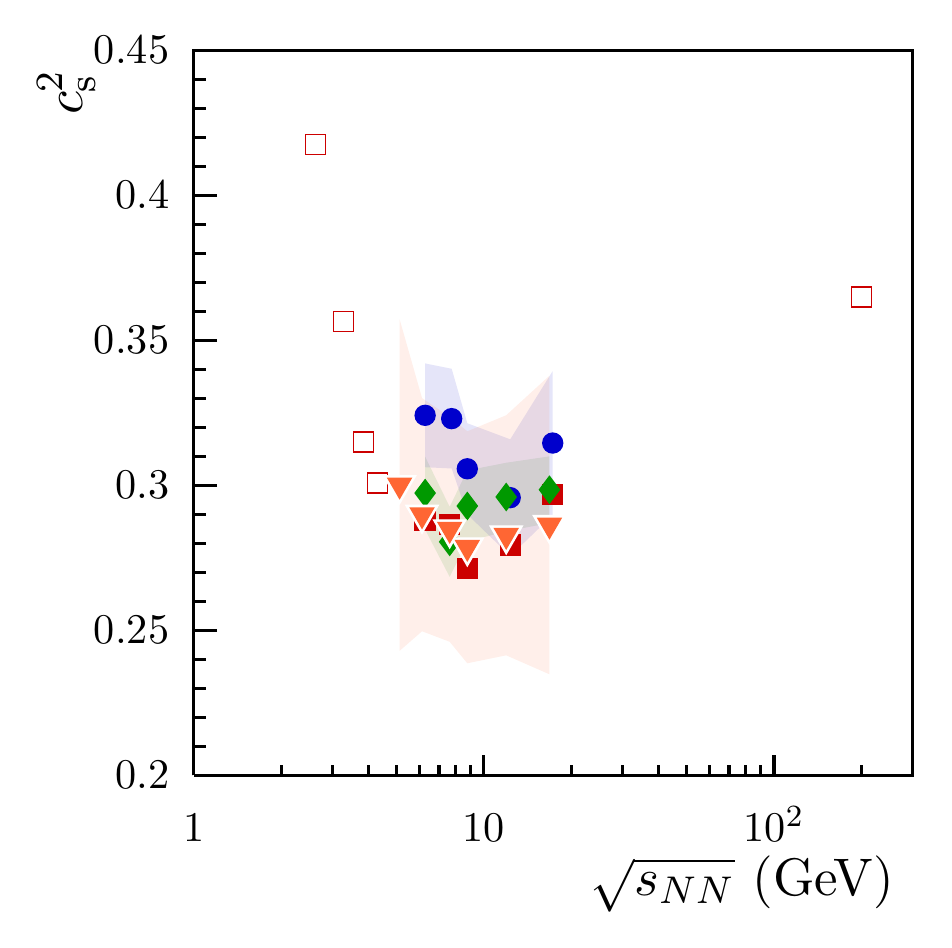}
 \put(72,32){\footnotesize World}
 \put(58,27){\footnotesize \textcolor{colorPbPb}{$\blacksquare$} Pb+Pb (NA49)}
 \put(58,22){\footnotesize \textcolor{colorAuAu}{$\square$} Au+Au (AGS)}
  \put(63,89){\footnotesize NA61/SHINE}
 \put(68,84){\footnotesize \textcolor{colorArSc}{$\blacktriangledown$} Ar+Sc}
 \put(68,79){\footnotesize \textcolor{colorBeBe}{$\blacklozenge$} Be+Be}
 \put(68,74){\textcolor{colorNN}{$\bullet$} {\footnotesize $N$+$N$}}
\end{overpic}
\caption{\textit{Left}: Mean pion multiplicity to the mean number of wounded nucleons versus the Fermi energy measure. \textit{Right}: The speed of sound $c_\text{s}^2$ as a function of center-of-mass collision energy as extracted from the data.}
\label{fig:kink_dale}
\end{figure}

The left panel of fig.~\ref{fig:kink_dale} presents the dependence of the ratio of the mean number of pions produced in a collision to the mean number of wounded nucleons $\langle\pi \rangle/\langle W \rangle$ versus the Fermi energy measure, $F=\left[\frac{(\sqrt{s_{NN}}-2m_{{N}})^3}{\sqrt{s_{{NN}}}}\right]^{1/4}\approx \sqrt[4]{s_{{NN}}}$. 
NA61/SHINE results: \textit{p+p}~\cite{pp_pions}, Be+Be~\cite{BeBe_pions} and Ar+Sc~\cite{ArSc_pions} were compared with NA49 results from Pb+Pb~\cite{PbPb_pions} and others. $\langle\pi \rangle/\langle W \rangle$ for Ar+Sc reactions equals that for \textit{N+N}\footnote{For \textit{p+p} interactions the figure shows isospin symmetrized values~\cite{pp_pions} marked as \textit{N+N}.} reactions at low SPS energies whereas it is consistent with that for central Pb+Pb reactions at high SPS energies. The behaviour of Ar+Sc stands in contradiction to Be+Be measurements, which are close to the Pb+Pb results except for the top SPS beam energy. The steepening of the slope of the $\langle\pi \rangle/\langle W \rangle$, which is considered a signal of the onset of deconfinement within the SMES~\cite{smes} is observed neither in \textit{central} Ar+Sc nor in \textit{central} Be+Be measurements.

The energy dependence of the sound velocities extracted from the data~\cite{pp_pions,BeBe_pions,ArSc_pions} are presented in fig.~\ref{fig:kink_dale}~(\textit{right}). Results from central Pb+Pb and Au+Au collisions exhibit a clear minimum (usually called the softest point) at $\sqrt{s_{NN}} \approx 10$ GeV consistent with the reported onset of deconfinement~\cite{PbPb_pions,PbPb_identified}. The energy range for results from Ar+Sc, Be+Be \textit{central} collisions and inelastic $p+p$ interactions is too limited to allow a significant conclusion about a possible minimum.

\subsection{Flow}
\begin{figure}[h]
  \includegraphics[width=0.47\textwidth]{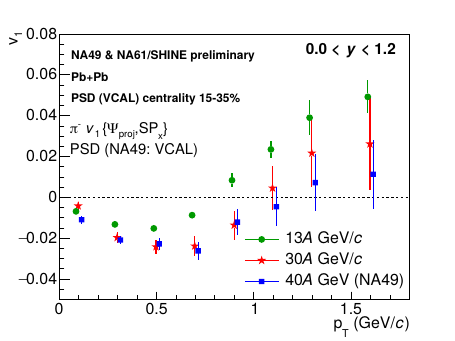}
 \includegraphics[width=0.47\textwidth]{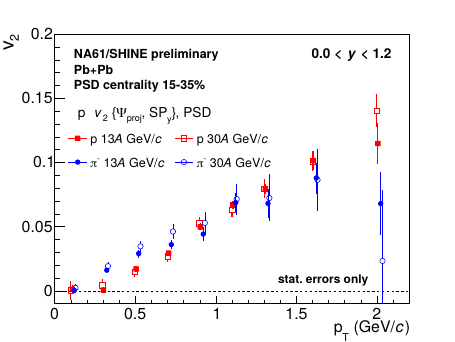}
 \caption{Preliminary results on negatively charged pions directed flow $v_1(p_T)$ and negatively charged pions and protons elliptic flow $v_2(p_T)$ for different collision energies.}
 \label{fig:flow}
\end{figure}

NA61/SHINE experiment reports results on anisotropic flow, measured in centrality selected Pb+Pb collisions at 13$A$ and 30$A$ GeV/$c$ beam momentum. The experimental setup allows tracking and particle identification over a wide rapidity range. Flow coefficients were measured relative to the spectator plane estimated with PSD, which is unique for NA61/SHINE. Preliminary results on the directed flow of negatively charged pion are presented in the left panel of fig.~\ref{fig:flow}.
In mid-central collisions the flow of negatively charged pions has minimum at $p_T \approx 0.2-0.6$ GeV/$c$. At the minimum the value $v_1$ is negative, while at $p_T \approx 0.7 - 1.3$ GeV/$c$ it changes sign to positive. In the region on sign change $v_1$ values measured at 13$A$ GeV/$c$ differs from the ones measured at 30$A$/40$A$ GeV/$c$. Dependence of $v_1(p_T)$ for protons produces in mid-central Pb+Pb collisions at 13$A$ GeV/$c$ beam momentum also reported. The $v_1(p_T)$ values for protons are positive and increase with transverse momentum until $p_T \approx 1.5 - 2.0$ GeV/$c$. Preliminary results on the centrality dependence of $dv_1/dy$ at mid-rapidity, measured in Pb+Pb collisions at 13$A$ GeV/$c$ were also reported. The slope of negatively charged pion $v_1$ is always negative. In contrast, the slope of proton $v_1$ changes sign for centrality of about 20$\%$. Preliminary results of elliptic flow, $v_2(p_T)$, for Pb+Pb collisions at 13$A$ and 30$A$ GeV/$c$ are presented in the right panel of fig.~\ref{fig:flow}. The $p_T$ dependence of $v_2$ coefficient of negatively charged pions differs from the one measured for protons, but it does not show any energy dependence.

\section{Search for critical point}
One of the main goals of the NA61/SHINE experiment is to locate the critical point (CP) of strongly interacting matter. The expected signal of a CP is a non-monotonic dependence of various fluctuations (e.g. multiplicity, transverse momentum, baryon density fluctuations). 

\begin{figure}[h]
\begin{minipage}{0.32\textwidth}
 \includegraphics[width=1.115\textwidth]{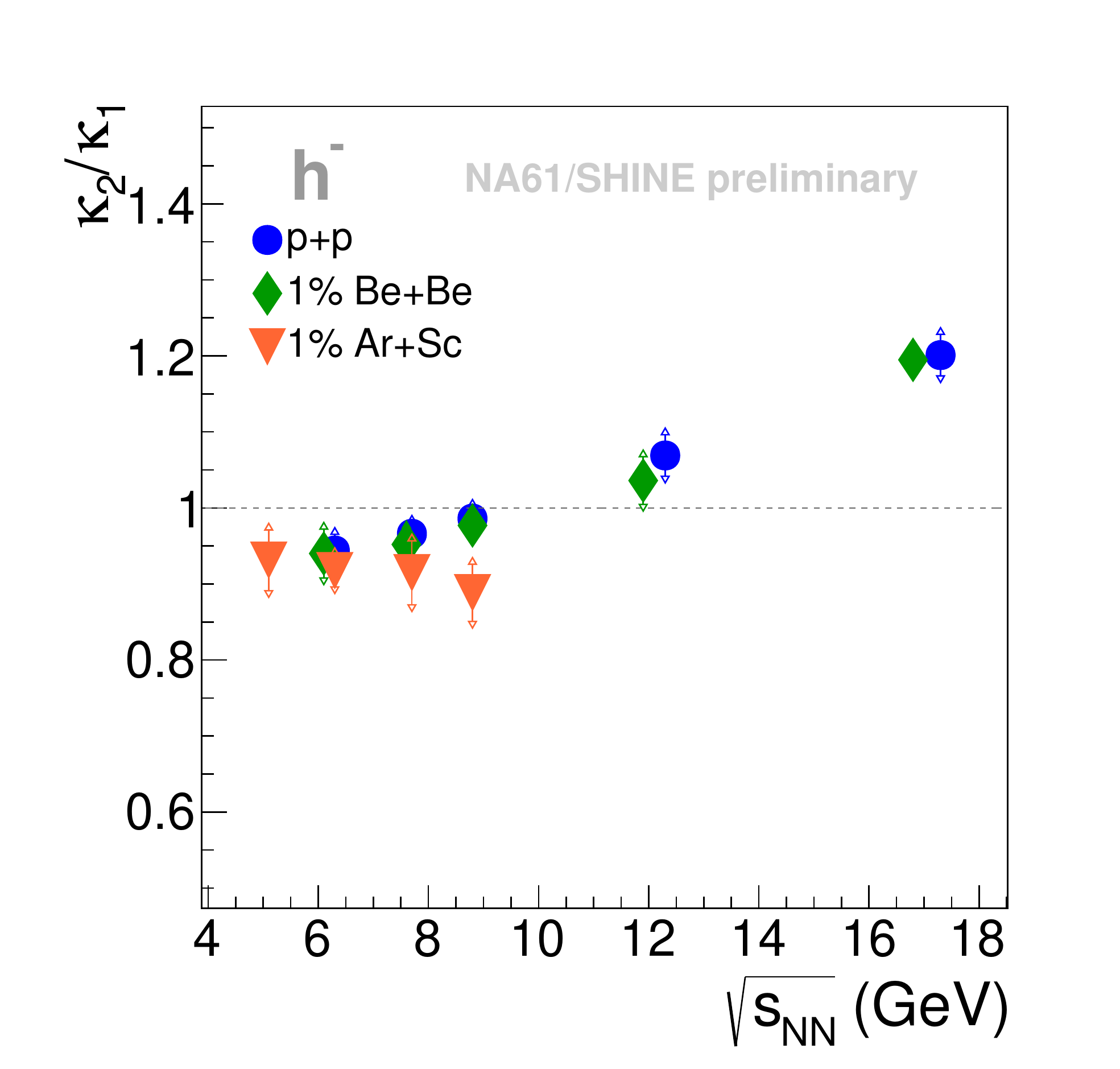}
\end{minipage}
\begin{minipage}{0.32\textwidth}
 \includegraphics[width=1.115\textwidth]{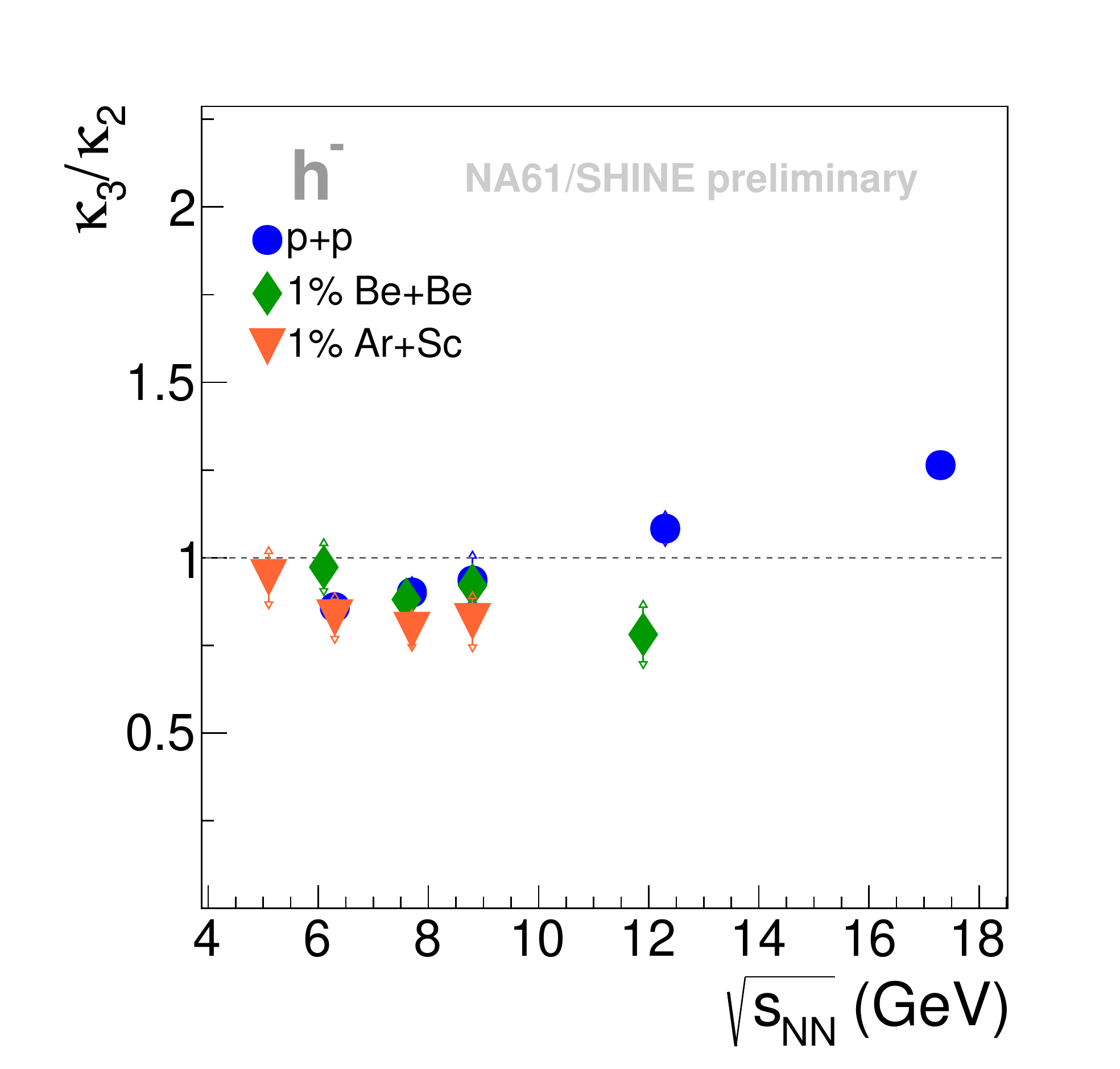}
\end{minipage}
\begin{minipage}{0.32\textwidth}
 \includegraphics[width=1.115\textwidth]{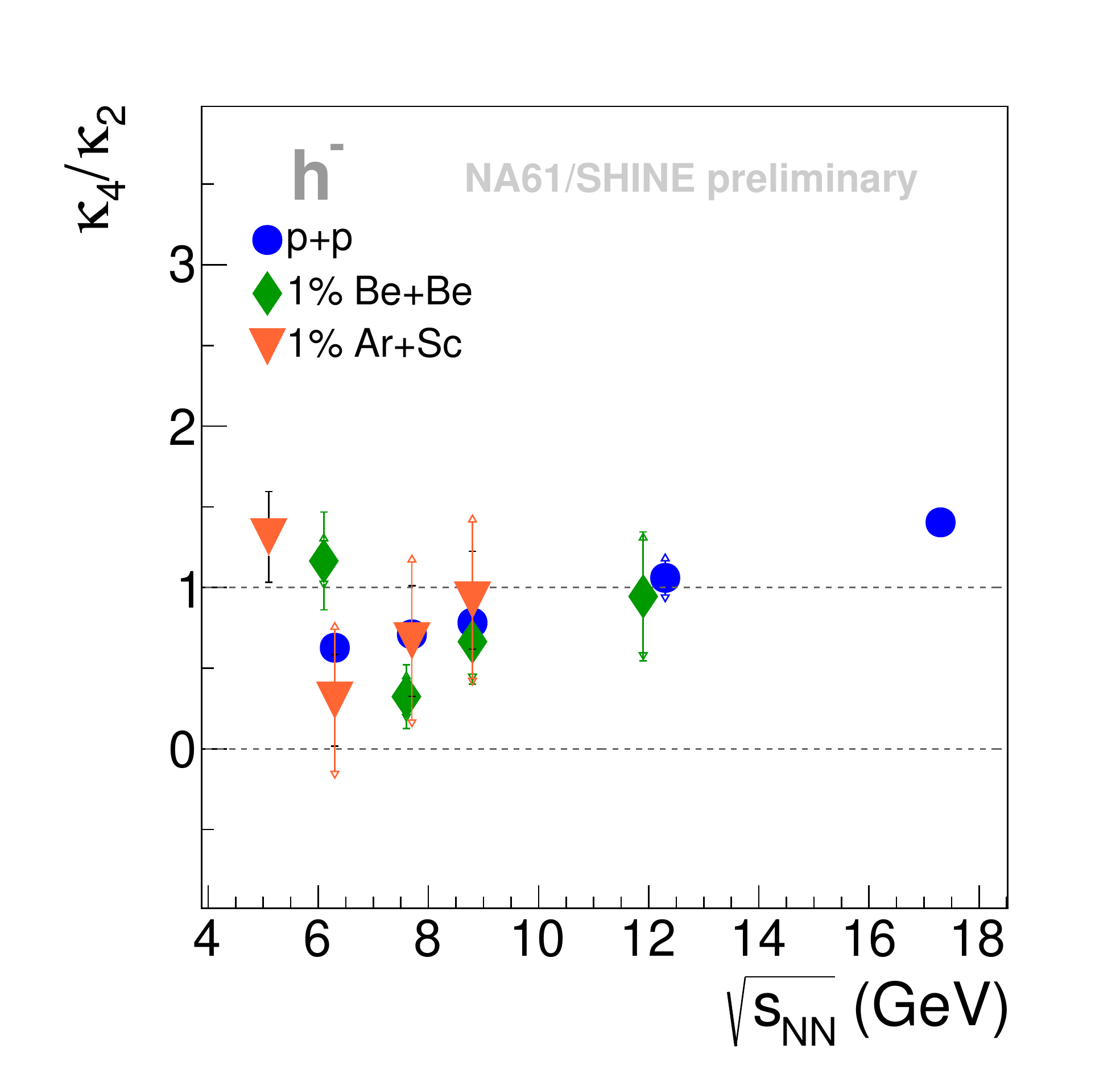}
\end{minipage}
 \caption{Preliminary results on the energy and system-size dependence of multiplicity fluctuations of negatively charged hadrons.}
 \label{fig:moments}
\end{figure}

To study the fluctuations in systems of different sizes one should use quantities insensitive to system volume (so-called intensive quantities), for example constructed by division of cumulants $\kappa_i$ of the measured distribution (up to fourth order), where $i$ is the order of the cumulant. For second, third and fourth order cumulants intensive quantities are defined as: $\kappa_2/\kappa_1$, $\kappa_3/\kappa_2$ and $\kappa_4/\kappa_2$. Fig.~\ref{fig:moments} shows energy dependence of second (\textit{left}), third (\textit{middle}) and fourth (\textit{right}) order cumulant ratio of negatively charged hadrons measured in \textit{p+p} interactions as well as Be+Be and Ar+Sc \textit{central} collisions.
The second-order cumulant ratio reveals an increasing difference between $\kappa_2/\kappa_1$ values from small systems (\textit{p+p} and Be+Be) and a heavier system (Ar+Sc) with collision energy. The \textit{p+p} results on third-order cumulant ratio tend to be higher than the ones from Be+Be at higher collision energies. Whereas, the fourth-order cumulant ratio indicates consistent values for all measured systems at given collision energy. More detailed studies are still needed, however, no structure indicating CP was observed so far.
 
\begin{figure}[h]
\centering
 \includegraphics[width=0.78\textwidth, trim={0 0.3cm 0 0}, clip]{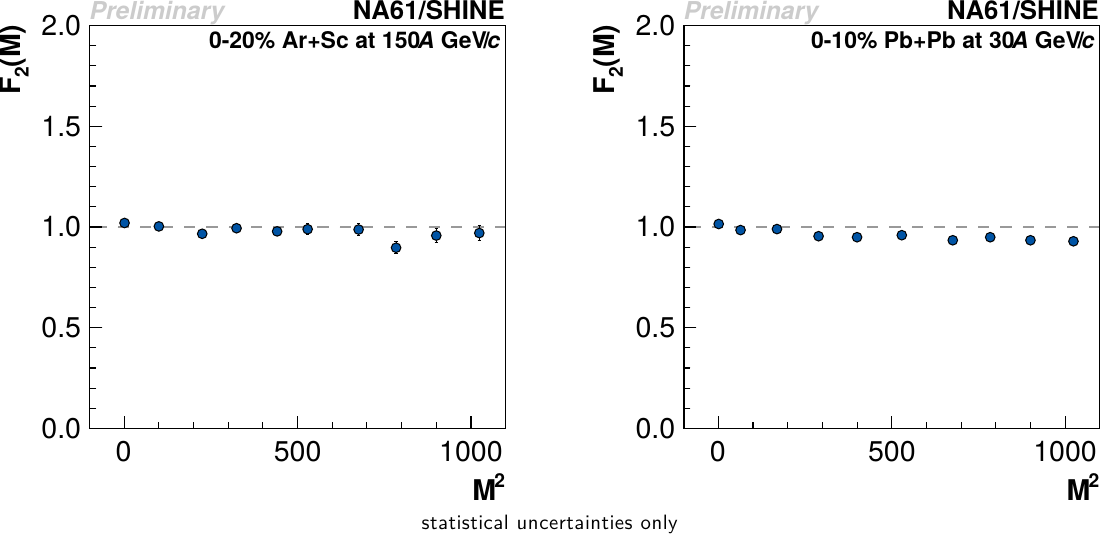}
 \caption{Preliminary results on $F_2(M)$ of mid-rapidity protons measured in 0-20$\%$ most central Ar+Sc collisions at 150$A$ GeV/$c$ (\textit{left}) and 0-10$\%$ most central Pb+Pb collisions (\textit{right}).}
 \label{fig:intermitency}
\end{figure}

Possible tool for search of CP is a proton intermittency. The proximity of CP is expected to manifest itself in local power-law fluctuations of the baryon density which can be searched for by studying the scaling behaviour of second factorial moments, $F_2(\delta)=\frac{\langle \frac{1}{M} \sum^M_{i=1} n_i(n_i-1) \rangle}{\langle \frac{1}{M} \sum^M_{i=1}n_i \rangle ^2}$, with the cell size or, equivalently, with the number of cells in ($p_x$ , $p_y$) space of protons at mid-rapidity~\cite{inter1,inter2,inter3}. NA61/SHINE measures $F_2(M)$ using using statistically independent points and cumulative variables. Preliminary results on $F_2(M)$ of mid-rapidity protons measured in 0-20$\%$ most \textit{central} Ar+Sc collisions at 150$A$ GeV/$c$ and 0-10$\%$ most central Pb+Pb collisions are presented in fig.~\ref{fig:intermitency} in left and right panels, respectively.
The intermittency index $\phi_2$ for a system freezing out at the QCD critical endpoint is expected to be $\phi_2 = 5/6$ assuming that the latter belongs to the 3-D Ising universality class. $F_2(M)$ of protons for Ar+Sc at 150$A$ GeV/$c$ and Pb+Pb at 30$A$ GeV/$c$ show no indication for
power-law increase with a bin size.

\section{Strangeness production in p+p interactions}
NA61/SHINE provides unique high precision results on strange hyperons production in \textit{p+p} interactions at 158 GeV/$c$: $\Xi^-$, $\overline{\Xi}^+$, $\Xi^0(1530)$ and $\overline{\Xi}^0(1530)$. The rapidity spectra were reported and compared to transport model predictions. Among tested models, EPOS describes the spectra of mentioned hyperons reasonably well. UrQMD agrees with $\Xi^-$ spectra measured by NA61/SHINE.
Additionally, the mean multiplicities of $\Xi^-$ and $\overline{\Xi}^+$ produced in inelastic p+p interactions at 158 GeV/c was used to quantify the strangeness enhancement in A+A collisions at the same centre-of-mass energy per nucleon pair. More details can be found in~\cite{pp_xi,pp_xi0}.

\section{Spectator-induced electromagnetic effects}
\begin{figure}[h]
 \includegraphics[width=1\textwidth]{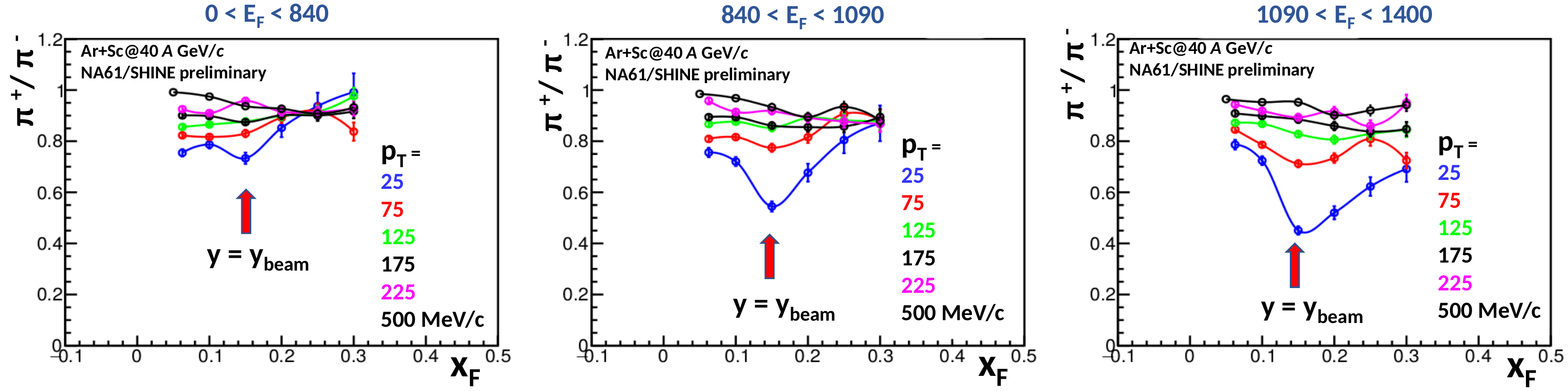}
 \caption{Dependence of $\pi^+/\pi^-$ ratio in the final state on $x_F$ measure and $p_T$ measured in PDS selected central (\textit{left}), semi-central (\textit{middle}) and semi-peripheral (\textit{right}) Ar+Sc at 40$A$ GeV/$c$ beam momentum.}
 \label{fig:electromagnetic}
\end{figure}

Charged spectators of the nuclear collision generate electromagnetic fields, which modify the trajectories of $\pi^+$ and $\pi^-$, which can result in charge splitting of directed flow. It is expected that the final state spectra $\pi^+$ and $\pi^-$ are modified by those fields. For convenience the double differential $\pi^+/\pi^-$ spectra ratios are studied~\cite{em1,em2}. NA61/SHINE reports new data on spectator-induced electromagnetic effects in Ar+Sc collisions at 40$A$ GeV/$c$ beam momentum. The results are presented in fig.~\ref{fig:electromagnetic} for PSD selected event centrality classes: central (\textit{left}), semi-central (\textit{middle}) and semi-peripheral (\textit{right}).
The magnitude of the $\pi^+/\pi^-$ spectra ratio modification increases with the peripherality of the collision. The effect is the strongest for pions moving at spectator velocity ($x_F\approx0.15$), close to spectator trajectory ($p_T\approx25$ MeV/$c$).

\section{Summary}
Presented experimental results on particle production properties reveal unexpected system-size dependence of $K^+/\pi^+$ ratio in mid-rapidity, which is not reproduced by any tested theoretical models. 
The velocity of sound extracted from the width of rapidity distribution of $\pi$ mesons produced in \textit{N+N} interactions and central Be+Be and Ar+Sc collisions is consistent with results for central Pb+Pb and Au+Au collisions. Measurements in a broader energy range are needed to conclude on a possible minimum of the sound velocity in small nuclei collisions. The results on the mean pion multiplicity to the number of wounded nucleons ratio and its collision energy dependence suggests an increase of the effective number of degrees of freedom already in \textit{central} Ar+Sc collisions at the top SPS energies. Moreover, presented results show no indications of the CP in Ar+Sc collisions at 150$A$ GeV/$c$. The measured directed flow of pions shows energy dependence, with the slope of negatively charged pions changing sign at different collision centralities in Pb+Pb collisions. Unique results on multi-strange baryons production in \textit{p+p} interactions at 158 GeV/$c$ beam momentum were discussed. Finally, the first observation of the electromagnetic effects in Ar+Sc at 40$A$ GeV/$c$ beam momentum was presented.

\end{document}